\documentstyle[twoside,epsfig]{article}

\begin{document}


\thispagestyle{empty}
\setcounter{page}{1}

\vspace*{0.88truein}

\centerline{\bf FINANCIAL FRICTION}
\vspace*{0.087truein}
\centerline{\bf AND MULTIPLICATIVE MARKOV MARKET GAMES}
\vspace*{0.37truein}
\baselineskip=10pt
\centerline{\footnotesize ERIK AURELL}
\vspace*{0.015truein}
\centerline{\footnotesize\it Matematiska Institutionen}
\baselineskip=10pt
\centerline{\footnotesize\it Stockholms Universitet,  SE-106 91 Stockholm, Sweden}
\baselineskip=10pt
\centerline{\footnotesize\it eaurell@matematik.su.se}
\vspace*{0.37truein}
\centerline{\footnotesize PAOLO MURATORE-GINANNESCHI}
\vspace*{0.015truein}
\centerline{\footnotesize\it CATS, The Niels Bohr Institute}
\baselineskip=10pt
\centerline{\footnotesize\it Blegdamsvej 17, DK-2100 Copenhagen, Denmark}
\baselineskip=10pt
\centerline{\footnotesize\it muratore@nbi.dk}
\baselineskip=10pt
\vspace*{10pt}

\vspace*{0.21truein}
\begin{abstract}
We study long-term growth-optimal strategies on a simple
market with transaction costs. We show that several problems of this sort can be 
solved in closed form, and explicit the non-analytic dependance of
optimal strategies and expected frictional losses of the friction
parameter.
\end{abstract}

\section{Introduction}	
\label{s:introduction}
\vspace*{-0.5pt}
\noindent
The purpose of this paper is to present results on optimal
control of investment strategies with friction. 
The discussion is for the most part couched in the frame-work
of growth-optimal portfolios of underlying securities,
the prices of which are smooth diffusions.
We will show how to treat such problems in a Fokker-Planck
framework, as an alternative to the Hamilton-Jacobi-Bellman equation methodology
used in financial economics~\cite{Merton,AkianMenaldiSulem}, and explicit that several problems
of this sort may be solved analytically in closed form.
It appears clearly in the Fokker-Planck formulation that optimal
strategies and expected friction lossses typically have a
singular dependence on the friction parameter. For underlying smooth
diffusions the friction losses scale with friction parameter to
the the fractional power of two thirds.
We consider also the more general situation, where the underlying is
not a smooth diffusion, generally limiting ourselves to the discrete
time case, and show how to relate to the optimal growth rate to the eigenvalue
of the proper evolution operator.
\\\\
The subject matter of this paper is not new, and the result that
optimal control of smooth diffusions
with linear friction is singular is well known in 
control theory, see e.g.~\cite{FlemingSoner}. 
We believe that juxtaposing derivations in terms of Fokker-Planck
and Hamilton-Jacobi-Bellman may help to make results obtained in the
second language more accessible to physicists, who are typically
more conversant with the first. On the other hand, the Fokker-Planck
approach has a certain simplicity, and facilitates order-of-magnitude
discussions and the use of a kind of dimensional analysis.
An original motivation for this work was to derive an option pricing
procedure by using growth-optimal criteria, generalizing work 
in the friction-less case~\cite{PaperA,PaperB}.
This is technically a harder problem. 
Various approximate solutions are possible and will be discussed in a forthcoming 
separate contribution~\cite{AurellMuratore_in_preparation}. 

\section{Stocks, bonds and transaction costs}
\label{s:stocks-bonds}
\noindent
In this section we derive the dynamics of a stock and
bond portfolio with proportional transaction costs in discrete
time, and then take the continuum limit.
We adhere to a representation of the dynamics given by a {\it pre-hedging} prescription.
Other representations are possible, and give rise
to dynamical equations, which although equivalent in content may
superficially look different.
A discussion of other prescriptions will be presented 
separately~\cite{AurellMuratore_in_preparation}. 
We start by assuming that a source of
of uncertainty is provided by the 
price movement of stock 
\begin{equation} 
S_{t+\Delta t}=u_t S_{t}
\label{eq:u_i-definition}
\end{equation} 
where time is discrete, $S_t$ is the price of a share  at time $t$ 
and the $u_t$'s are independent, identically distributed
random variables.
We also assume a risk-less security (a bond or a bank account), with
price movement
\begin{equation} 
B_{t+\Delta t}=r B_{t}\,\, .
\label{eq:r-definition}
\end{equation}
By a change of numeraire we can always take $r$ equal to one.
We then assume proportional transaction costs. That is, an investor who holds
some capital $C$ in stock or bonds at some time $t$, and would like to change
his holding to the other security, would after the operation hold  
capital $(1-\gamma)C$, where $\gamma$ is a non-zero number.  
Several interpretations of $\gamma$ are possible. One is simply that there
are direct fees coupled to every transaction. In real markets there are various
fees, some proportional as we assume, some a fixed commission per transaction,
and some with other structure such as different proportional commissions for
buying and selling stock. For a real operator on a real market there are also
indirect costs, such as overhead and operating costs, which, insofar as
they can be determined from the outside, are perhaps roughly proportional to trading
volume. 
Another interpretation of $\gamma$ is that there is really no one price,
but only bid and ask prices, with some spread in between.
If spread is proportional to price this will have the same
effect as proportional friction.
\\\\
We consider at time $t$ an investor who holds
capital $W_{t}$, of which $\Delta W^S_{t}$ is invested in stock. The fraction invested in
stock is hence $l_{t}=\frac{\Delta W^S_{t}}{W_{t}}$.
The investor would like to change the amount invested
in stock to $\alpha_t W_{t}$, and when
rebalancing his portfolio he has to pay
transaction costs of
\begin{equation}
\Delta W^{TC}_{t} = \gamma |\alpha_tW_t-l_{t}W_t|
\end{equation}
The total wealth of the investor before the market moves again is hence
$\left(1-\gamma |\alpha_t-l_{t}\right)W_{t}$,
and after the market has moved at time $(t+\Delta t)$ the wealth and fraction of
wealth invested in stock are
\begin{eqnarray}
W_{(t+\Delta t)} &=&\left(u_{t+\Delta t}\alpha_t +(1-\alpha_t)-\gamma |\alpha_t-l_{t}|
\right)W_{t} 
\label{eq:W-pre-prescription-dynamics}\\
l_{(t+\Delta t)} &=& \frac{u_{t+\Delta t}\alpha_t}{u_{(t+\Delta t )}\alpha_t+(1-\alpha_t)
-\gamma |\alpha_t-l_{t}|}
\label{eq:l-pre-prescription-dynamics}
\end{eqnarray}
We see that the $l_{t}$'s can be considered as consecutive
positions in a series of gambles, or in a repeated game against Nature. 
The control variables $\alpha_{t}$ are on the other hand strategy choices
made at different times (and in different positions) in the game.
The game is multiplicative, since pay-off is proportional to capital,
and it is Markov because the new capital and new position 
($W_{(t+\Delta t)}$ and $l_{(t+\Delta t)}$) depend parametrically on the
previous position ($l_t$).
A simple special case 
is when transaction costs
are zero and the 
dependence of wealth on previous position drops out. The game is then Bernoulli, and
the only relevant variable is the strategy $\alpha_t$. 
The optimal strategies according to the growth-optimal criterion
in this class of models were determined by Kelly~\cite{Kelly}. For recent 
papers and reviews with references to the later
literature, see e.g.~\cite{HakansonZiemba,MaslovZhang,PaperA,PaperB}.   
\\\\
The continuum limit of (\ref{eq:W-pre-prescription-dynamics})
and (\ref{eq:l-pre-prescription-dynamics})
is reached by taking the random events to occur
at times $0,dt,2dt,\ldots$, and modellling the stock price fluctuations 
by a stochastic differential
\begin{equation}
u_{t}=1+\mu dt +\sigma dB_t
\label{market}
\end{equation}
$dB_t$ being increments of a Wiener process. Furthermore we assume that the
control variable $\alpha_t$ is smoothly related to the position $l_{t}$ by
\begin{equation}
\alpha_{t}=l_{t}+\beta_t dt
\label{control}
\end{equation}
The wealth fraction then changes as
\begin{eqnarray}
l_{(t+dt)}
&=& l_{t} + [(\mu -\sigma^2l_{t})l_{t}(1- l_{t}) +\beta_t +\gamma l_t|\beta_t|]dt +
\nonumber \\
&&\quad [\sigma l_{t}(1- l_{t})  ]dB_{t} \,+\,\hbox{h.o.t}
\end{eqnarray}
where we have taken $dB_t$ to be formally of size $\sqrt{dt}$ and kept all terms
up to linear in $dt$.
Hence
\begin{eqnarray}
dl_{t}&=&\left((\mu -\sigma^2l_{t})l_{t}(1- l_{t}) +\beta_t +\gamma l_t |\beta_t|\right)dt +
 \left(\sigma l_t(1- l_{t})  \right)dB_{t}
\label{eq:l-SDE}
\end{eqnarray}
\noindent
The wealth process can be written down in a similar fashion, after applying
Ito's lemma in a form  equivalent to
\begin{eqnarray}
d\log W_{t} &=&\left(\mu l_{t}-\frac{\sigma^2}{2}l^2_{t} -\gamma |\beta_t|\right)dt + \sigma l_{t} dB_{t}
\label{eq:logW-SDE}
\end{eqnarray}
It is also useful for further reference to observe that without control
the stochastic differential
equation (\ref{eq:l-SDE}) preserves the three intervals 
$[-\infty,0]$, $[0,1]$ and $[1,\infty]$. A candidate for an optimal control
will not move the process out of the best of the three intervals.
Assuming for definiteness  that the best interval is $[0,1]$ we
introduce the variable $\xi = \log\frac{l}{1-l}$ and find
\begin{equation}
d\xi_{t}=[\kappa + \mu -\frac{1}{2}\sigma^2]dt + \sigma dB_{t}
\qquad \kappa = (1+e^{\xi})\left(\beta (1+e^{-\xi}) +\gamma |\beta_t|\right)
\label{eq:xsi-SDE}
\end{equation}
The variable $\xi$ ranges from $-\infty$ to $+\infty$. Without control
($\beta$ and $\kappa$ equal to zero) $\xi$ follows a normal diffusion
process with constant drift equal to $\mu -\frac{1}{2}\sigma^2$. 
For long times without control the process therefore
ends up with $\xi$ far from the origin, that is $l$ close
to zero or one. If the best intervals in $l$ are $[-\infty,0]$ or
$[1,\infty]$, the process without control in an analogous manner
ends up in $0$ or $1$, respectively. In other words, if you do
not rehedge you eventually end with all your money in stock,
or, depending on the parameters, with all your money in bonds.

\section{Long-term growth with transaction costs}
\label{s:long-term-growth-optimal}
\noindent
We start with some general remarks, borrowing
some of the terminology from the theory of dynamical systems.
Let us consider the dynamics of wealth and wealth fraction, where we write the
strategy ($\alpha_t$) as a function of position ($l_t$):
\begin{eqnarray}
W_{(t+\Delta t)} &=&\left(u_{(t+\Delta t)}
\alpha_t(l_t) +(1-\alpha_t(l_t))-\gamma |\alpha_t(l_t)-l_{t}|\right)W_{t} \\
l_{(t+\Delta t)} &=& \frac{u_{(t+\Delta t)}\alpha_t(l_t)}{u_{(t+\Delta t)}
\alpha_t(l_t)+(1-\alpha_t(l_t))
-\gamma |\alpha_t(l_t)-l_{t}|}
\end{eqnarray}
The dynamics in $l$ induces a linear evolution
equation for the probability distribution over the states
in the corresponding Markov chain (i.e., the values of $l$). The asymptotic distribution
in $l$ is the eigenvector corresponding to the highest eigenvalue of the corresponding
operator, let that be written $P_0(l;[\alpha])dl$.
The long-term growth rate is thus
\begin{eqnarray}
\lambda(\gamma;[\alpha]) &=& \int P_0(l;[\alpha])\,E[ \log(1+\alpha(l)(u-1)
-\gamma |\alpha(l)-l|)]\, dl
\label{eq:lambda-as-function-of-alpha}
\end{eqnarray}
the expected logarithmic return with respect
to the joint distribution over $l$ and the random variable $u$
\footnote{In our model $l$ and $u$ are independent, so
the joint distribution is the product of the distributions
over $l$ and $u$ separately. In a proper game one would expect that
the $u$ would not be independent of $l$ -- and also dependent on
the positions of the other players in the game.}.
Maximizing long-term growth  is a non-trivial
problem because the strategy $\alpha$ enters both in the function
to be averaged, i.e. $\log(1+\alpha(l)(u-1)
-\gamma |\alpha(l)-l|)$, and in the measure of integration. 
\\\\
To proceed further we turn to the continuum limit, as defined by equations
(\ref{eq:l-SDE}) and (\ref{eq:logW-SDE}). We will consider a model
problem, and show how that can be solved by Fokker-Planck methods. Then
we will repeat the model problem in the Hamilton-Jacobi-Bellman
formalism, and then return to the full problem of equations 
(\ref{eq:l-SDE}) and (\ref{eq:logW-SDE}).
A kind of dimensional analysis and possible extentions of the procedure
are discussed separately in section~\ref{s:dimension-analysis}.

\subsection{A model problem}
\label{ss:model-problem}
\noindent
Let us consider the stochastic differential equation
\begin{equation}
dx_{t}=-\left[\frac{\partial V}{\partial x}\right] dt+\sqrt{2D}\, dB_t
\label{eq:model-problem-SDE}
\end{equation}
and suppose that we wish to maximize
\begin{equation}
U=E\,\left[ax -\frac{1}{2}bx^2 - c\, |\frac{\partial V}{\partial x}|\right]
\label{eq:functional-to-maximize}
\end{equation}
where the expectation value is taken with respect to the stationary
distribution of $x$,
\begin{equation}
E\,\left[f(x)\right]=\int N^{-1}e^{-\frac{V(x)}{D}}f(x)dx.
\end{equation}
The normalization factor $N$ is fixed by $E[1]=1$.
The problem is hence to maximize with Lagrange multiplier $q$ the
functional
 \begin{equation}
Q=N^{-1} \int e^{-\frac{V(x)}{D}}\left(
ax -\frac{1}{2}bx^2 - c\, |\frac{\partial V}{\partial x}|\right) dx
+ q \left(\int e^{-\frac{V(x)}{D}}dx -N\right)
\end{equation}
We vary with respect to $V'(x)$ to find
 \begin{eqnarray}
\frac{\delta Q}{\delta V'(x)}
&=&  N^{-1} \int_x^{\infty} e^{-\frac{V(x)}{D}}(-\frac{1}{D})\left(
ax -\frac{1}{2}bx^2 - c\, |\frac{\partial V}{\partial x}| +qN\right) dx
\nonumber \\
&& - N^{-1} c\, \hbox{sign}(\frac{\partial V}{\partial x})
\label{eq:Q-first-variation}
\end{eqnarray}
At an extremum we have $\frac{\delta Q}{\delta V'(x)}=0$, for every $x$ such
that $V(x)<\infty$. If we differentiate with respect
to $x$ two terms cancel, and we have the simpler equation
\begin{equation}
ax -\frac{1}{2}bx^2 +qN-2Dc\,\delta(V'(x))V''(x) =0.
\label{eq:Q-first-variation-differential}
\end{equation}
Clearly it is not possible to satisfy~(\ref{eq:Q-first-variation-differential})
unless $V'=0$. If $V'=0$ the equation can be understood in the distributional
sense, for instance by broadening the $\delta$ function to be $1/\epsilon$
on the interval $[-\epsilon/2,\epsilon/2]$, and zero elsewhere.
Then  
(\ref{eq:Q-first-variation-differential}) is equivalent to the ordinary differential
equation $V''=\frac{\epsilon}{2Dc}(ax -\frac{1}{2}bx^2 +qN)$ as long as $V'$, the
integral of $V''$, is not larger than $\epsilon/2$ in absolute value. This equation
can obviously be satisfied on an interval in $x$ of finite width.
To find the optimal interval without making an explicit assumption on the regularization,
we just assume
that $V$ is large outside an interval $[x_-,x_+]$ and that $V'$
is small in the same interval. The quantity to maximize can then be
made arbitrary close to 
\begin{equation}
U(x_-,x_+) = \frac{a}{2}
\frac{(x_+)^2-(x_-)^2}{x_+-x_-}
 -\frac{b}{6}\frac{(x_+)^3-(x_-)^3}{x_+-x_-}
-\frac{2Dc}{x_+-x_-}
\label{eq:x-functional}
\end{equation}
which is maximized by
\begin{equation}
x_{\pm} = \frac{a}{b}\,\pm\,
\frac{1}{2} \left(\frac{24cD}{b} \right)^{\frac{1}{3}}
\label{eq:x-maximum}
\end{equation}
The optimum control is thus given by a ``deep-pit'' potential with mid-point
$\frac{a}{b}$ and width $\left(\frac{24cD}{b}\right)^{\frac{1}{3}}$.
The value of the maximum is
\begin{equation}
U_{\max} = \frac{a^2}{2b} - \frac{1}{2}\left(9\, b\,D^2\,c^2\right)^{\frac{1}{3}}
\label{eq:U-maximum}
\end{equation}
The loss of utility is therefore proportional to friction parameter $c$ to
the power two thirds.

\subsection{Hamilton-Jacobi-Bellman}
\label{ss:HJB}
\noindent
Dynamical programming is formulated in the setting that we have a SDE
\begin{equation}
\begin{array}{lll}
d\xi_t=f(\xi_t,u)\,dt+
\sigma(\xi_t,u)\,dB_t \,, & \quad
 &  \xi_s=x
\end{array}
\label{eq:controlled-SDE}
\end{equation}
where $u=u(x,t)$ is a dynamic control function. We wish to extremize
a functional, generally of a form like
\begin{equation}
J(x,s)=\inf_{u \in \Re} \left\{ E_{\xi_s=x}\int_{s}^{T}dt\,L(\xi_t,D^{+}\xi_t)+
J(\xi_T,T)\right \}
\label{eq:functional}
\end{equation}
where the forward derivative is defined as
\begin{equation}
D^+F(\xi_t) = \lim_{h \downarrow 0} E_{\xi_s=x}\left\{\frac{F(\xi_{t+h},t+h)-F(x,t)}{h}\right\}.
\label{eq:forward-derivative}
\end{equation}
The solutions of 
(\ref{eq:controlled-SDE}) 
are assumed to be 
smooth diffusions for any admissible choice of $u$.
In our model problem there is no control of the diffusion term, i.e.
$\sigma(\xi_t,u)=\sqrt{2D}$ does not depend on $u$. The drift term
$f(\xi_t,u)$ is on the other hand $-\frac{\partial V}{\partial x}$,
and we may also identify 
$D^+\xi_t$ with $f$.
Hence, if $L(\xi_t,D^{+}\xi_t)=-a\xi_t+\frac{b}{2}\xi_t^2 +c|D^+\xi_t|$
and $J(\xi_T,T)=0$,
then $\frac{J(x,s)}{T-s}$ is the minimum of the average value of the functional $L$
which can be achieved in the time interval $[s,T]$, given that the process
starts in point $x$ at time $s$. If the process
(\ref{eq:controlled-SDE}) is stationary, then the large $T$ limit of this average
will be almost surely opposite in sign but equal in absolute
value of the expectation value to be maximized in
(\ref{eq:functional-to-maximize}). 
\\\\
The idea is to recast the problem in terms of a differential equation.
Bellman's principle states that for any $h$ between $0$ and $T-s$
\begin{equation}
J(x,s)=
\inf_{u \in \Re} 
\left\{ E_{\xi_s=x}\int_{s}^{s+h}dt\,L(\xi_t,D^{+}\xi_t)+
J(\xi_{s+h},s+h)\right \}
\label{eq:functional_s_to_s+h}
\end{equation}
With the assumptions that the process
(\ref{eq:controlled-SDE}) is a smooth diffusion (already made), and that $J(x,s)$
is a smooth function of $x$ and $s$ (which follows if 
(\ref{eq:controlled-SDE}) is a smooth diffusion and $s$ less than $T$),
we have
\begin{eqnarray}
0&=&\lim_{h \downarrow 0}
\inf_{u \in \Re} 
E_{\xi_s=x}\left\{ \frac{1}{h}\,\int_{s}^{s+h}dt\,
L(\xi_t,D^{+}\xi_t)+\frac{J(\xi_{s+h},s+h)-J(x,s)}{h}\right\}=
\nonumber\\
&&=
\inf_{u \in \Re} \left\{ L(x,u) + \partial_{s}+f(x,u)\partial_{x}+\frac{\sigma(x,u)^2}{2}\partial_{x}^{2} 
\right\}\,J(x,s)
\label{eq:differential_dynamic_programming}
\end{eqnarray}
Substituting $L$, $\sigma(\xi_t,u)$ and introducing $\lambda=-J$ we have
\begin{eqnarray}
&&\partial_{s}\lambda+\sup_{f}\,\{f\,\partial_{x}\lambda+
D\,\partial_{x}^{2}\lambda 
+ a x -\frac{b}{2}-c |f|\}=0
\nonumber\\
&&\lambda(x,T;T)=0
\label{eq:modelhjb}
\end{eqnarray}
Naive variation with respect to $f$ leads to $\partial_x\lambda = c\,\hbox{sign}(f)$.
This is problematical because the sign function is not invertable, and we therefore
have no equation for the control. Furthermore, control 
entirely drops out of (\ref{eq:modelhjb}), and we have instead
the much simpler equation
$\partial_{s}\lambda+
D\,\partial_{x}^{2}\lambda 
+a x -\frac{b}{2}=0$, with solution $\lambda(x,s)=(T-s)(ax-\frac{b}{2}x^2)-\frac{Db(T-s)^2}{2}$.
The average $\lambda/(T-s)$ 
decreases asymptotically linearly as $-Db(T-s)/2$ with increasing $T$.
In fact, all we have solved for is the expected value of $ax-\frac{b}{2}x^2$ of an uncontrolled
random walk with diffusion coefficient $D$ starting from $x$ and moving for a time
$T-s$. 
\\\\
The problem is that motion in a ``deep-pit'' potential, which we already know is the asymptotically
optimal control, is not a smooth diffusion at the boundary. 
For a full discussion, see e.g.~\cite{FlemingSoner}, to extend (slightly)
the Hamilton-Jacobi-Bellman in a less rigorous manner, assume an
interval $[x_-,x_+]$ such that the restoring force $f$
goes to positive and negative infinity, respectively, when $x$ tends to the boundary points
$x_-$ and $x_+$ from the inside. Outside the interval $f$ is supposed to remain infinite,
so that the process is constrained to be in $[x_-,x_+]$.
If further the variational equation
$\partial_x\lambda = c\,\hbox{sign}(f)$ is interpreted in the distributional sense,
for instance by broading the sign function to $\tanh\frac{f}{\epsilon D}$, then $f$
is determined as a function of $\partial_x\lambda$ inside the interval, of size $\epsilon$ away
from the boundaries, and 
$\lambda$ satisfies the boundary conditions
\begin{equation}
\partial_{x} \lambda|_{x=x_-}\, =\,
\left(-\partial_{x} \lambda|_{x=x_+}\right)\,=\,c
\label{boundariesmodel}
\end{equation}
The dynamic programming equation (\ref{eq:modelhjb})
can now be solved for large $T$, if we make the 
ansatz 
\begin{eqnarray}
\lambda(x,s;T)\rightarrow \frac{(T-s)}{\tau} + \mbox{asymptotically constant in $T$} 
\label{Oseledec}
\end{eqnarray}
When (\ref{Oseledec}) holds true 
one integration in $x$ in (\ref{eq:modelhjb}) gives
$
D\partial_{x} \lambda=A + 
\frac{1}{\tau}x -\frac{a}
{2}x^2 +\frac{b x^{3}}{6}
$
and the two boundary conditions read
\begin{equation}
A +\frac{1}{\tau}x_- -\frac{a}{2}x_-^2+\frac{b}{6}x_-^3 \,=\,
-\left(A +\frac{1}{\tau}x_+ -\frac{a}{2}x_+^2+\frac{b}{6}x_+^3\right)
\,=\,Dc
\end{equation}
which are easily separated into one equation for the constant of
integration $A$, and one equation for $\tau$, which is identical
to (\ref{eq:x-functional}). We have therefore derived the same
result again in the HJB formalism.
As a consistency check for the solution one verifies that in the interval 
$[x_-,x_+]$ the absolute value of the gradient of $\lambda$ is bounded by $c$.

\subsection{The full stock and bond problem}
\label{ss:stock-and-bond}
\noindent
It is convenient to consider the problem in terms of the transformed variable
$\xi$ of (\ref{eq:xsi-SDE}), which obeys a stochastic differential equation
entirely similar to the model problem (\ref{eq:model-problem-SDE})
with $-\partial_x V = (\kappa +\mu-\sigma^2/2)$
and $D=\sigma^2/2$, but where
the function to maximize is
\begin{equation}
F=E\,\left[\mu
\left(\frac{e^{\xi}}{1+e^{\xi}}\right)
-\frac{1}{2}\sigma^2
\left(\frac{e^{\xi}}{(1+e^{\xi})^2}\right)
-
\frac{e^{\xi}}{1+e^{\xi}}\frac{\gamma|\kappa|}{1+e^{\xi}(1+\gamma\,\hbox{sign}(\kappa))}\right]
\label{eq:F-functional-to-maximize}
\end{equation}
In deriving (\ref{eq:F-functional-to-maximize}) we have expressed the
control $\beta$ (of $l$) in terms of the control $\kappa$ (of $\xi$).
Let us write $F=E[g(x,u)|_{u=\partial_x V}]$. The equation analogous to
(\ref{eq:Q-first-variation-differential}) will then be
\begin{equation}
\frac{\partial^2 V}{\partial x^2}\frac{d^2g}{du^2}
|_{u=\partial_x V}
-\frac{1}{D} \frac{\partial V}{\partial x} \frac{dg}{du}|_{u=\partial_x V}
+\frac{d^2g}{dxdu}|_{u=\partial_x V}
+\frac{1}{D}(g-qN) = 0
\end{equation}
This equation will not have solutions (when $V<\infty$) unless $\kappa$
vanishes. The invariant density is hence
\begin{equation}
P(\xi) = N^{-1} \exp(\xi\frac{\mu-\sigma^2/2}{\sigma^2/2})\qquad \xi\in [\xi_-,\xi_+]
\end{equation}
and the two endpoints $\xi_-$ and $\xi_+$ can be determined by maximizing
\begin{eqnarray}
F(\xi_-,\xi_+)&=& N^{-1}\int_{\xi_-}^{\xi_+}
				e^{(\frac{2\mu}{\sigma^2}-1)\xi}
				\left(
					\mu\left(\frac{e^{\xi}}{1+e^{\xi}}\right)
					-\frac{1}{2}\sigma^2 \left(\frac{e^{\xi}}{(1+e^{\xi})^2}\right)
				\right)
			d\xi	\nonumber \\
		&&-\gamma\frac{\sigma^2}{2}\frac{1}{N}
			\left(
				e^{(\frac{2\mu}{\sigma^2}-1)\xi_-}
					\frac{e^{\xi_-}}{1+e^{\xi_-}} \frac{1}{1+e^{\xi_-}(1+\gamma)}
			 \right)
		\nonumber \\
		&&\quad-\gamma\frac{\sigma^2}{2}\frac{1}{N}
			\left(
				e^{(\frac{2\mu}{\sigma^2}-1)\xi_+}
					\frac{e^{\xi_+}}{1+e^{\xi_+}} \frac{1}{1+e^{\xi_+}(1-\gamma)}
			\right)
\label{eq:F-functional-of-boundary-points}
\end{eqnarray}
In the small $\gamma$ limit it is sufficient to observe that the optimal interval
is centered on 
$l=\frac{\mu}{\sigma^2}$, 
and that the width is determined by
(\ref{eq:x-maximum}) with $D=\frac{\sigma^2}{2}
(\frac{\mu}{\sigma^2})^2(1-
\frac{\mu}{\sigma^2})^2$, $b=\sigma^2$ and $c=\gamma$. Inserting these quantities
we have
\begin{equation}
\Delta l \approx \left(12\gamma\, (\frac{\mu}{\sigma^2})^2(1-
\frac{\mu}{\sigma^2})^2 \right)^{\frac{1}{3}} 
\qquad\hbox{$\gamma$ small}
\end{equation}
When $\mu/\sigma^2$ is close to zero or one
 the width of the holding interval vanishes. This is
natural, since if the optimal investment is to hold all money in stock 
or bonds no rehedging is necessary. On the other hand,
when $\mu/\sigma^2$ is around one half the optimal holding interval is
about $0.9\,\gamma^{\frac{1}{3}}$. For realistic values of $\gamma$ this is not
a small number. For instance, if $\gamma$ is one percent, and the best portfolio
without friction is to hold half each in stock and bonds,
then the optimal strategy is to not rehedge unless the fraction in stock
falls below 40 or rises above 60 percent. 

\section{Dimensional analysis}
\label{s:dimension-analysis}
\noindent
It is instructive to see that most of the results can be derived
by a kind of dimensional analysis. Let us consider the model problem,
and call the dimension of the quantity to maximize $U$. In the full
problem it would be wealth per time. Let $L$ be the dimension
of the quantity $x$.  The dimensions of the parameters are then
\begin{equation}
[\,a\,] = \frac{U}{L}\quad [\,b\,]=\frac{U}{L^2}\quad [\,c\,]=\frac{UT}{L}\quad
[\,D\,] = \frac{L^2}{T}
\end{equation}
From $a$ and $b$ we can form one quantity with dimension $L$, namely $a/b$.
This is the optimal value of $x$ without friction. From
$b$, $D$ and $c$ we can also form one quantity with dimension $L$,
$(Dc/b)^{\frac{1}{3}}$, which, up the numerical prefactor, is
indeed the width of the holding interval. 
\\\\
From $D$ and the width of the holding interval ($d$) we can form a characteristic
time between rehedgings $\tau=d^2/D$. If the growth-optimal criterion is to have
a special significance for an individual investor we must suppose that his investment
horizon is much longer than $\tau$. For realistic parameter values this implies
surprisingly long time scales. If we use the previously
discussed example with $\gamma=0.01$, and assume $\sigma$ one percent per day,
then the characteristic time is 800 days, about three years of trading time.
If we assume instead the extreme and probably unrealistic value of $\gamma$ one
part in a million, then the holding interval is still one percent in width, and
the characteristic time is measured in days.
The conclusion is, that in the very simple model of one stock and one bond,
with transaction costs, it is delicate and rather doubtful if times are ever
sufficiently long that the growth-optimal criterion is unquestionably valid.
If we enlarge the model to include more securities, especially derivative
securities the return distributions of which depend explicitly on time to
maturity, this conclusion can only be strengthened.  For the growth-optimal criterion
to be of special sig\-nifi\-cance one would have to discuss not individual
investors, but evolving populations of investment strategies, and to show,
if true, that growth-optimal strategies in multiplicative games
approximate non-cooperative Nash
equilibria, selected by the Nash ``mass-action'' principle
\cite{Nash-PhD}.
For a recent discussion of populations of evolving strategies, see also \cite{Farmer}.
\\\\
We can also use dimensional analysis to consider sources of uncertainty other
than smooth diffusions. If a noise term $dw_t$ is formally taken to be $(dt)^{\alpha}$,
then $D$ will be of dimension $L^2/T^{2\alpha}$, the characteristic width
$d=c^{\frac{\alpha}{1+\alpha}}D^{\frac{1}{2+2\alpha}}b^{-\frac{\alpha}{1+\alpha}}$
and $\tau=\left(\frac{d^2}{D}\right)^{\frac{1}{2\alpha}}$. 
The expected friction losses per unit time are $cd/\tau$. A simple strategy with this
scaling is to bring back the portfolio to the optimum (cost $cd$) every time the fraction
goes out of a band of width $d$ (event occuring with frequency $1/\tau$).
For smooth diffusion this is the same scaling as in~(\ref{eq:U-maximum});
only the prefactor will be better with the optimal strategy.
Since for realistic parameter
values the characteristic times are typically long, 
it however probably suffices that the noise is
Gaussian and uncorrelated
on those time scales for the optimal control of smooth diffusions that
have mainly been discussed here to be relevant.

\section*{Acknowledgments}
\noindent
We thank 
P.~Dimon, 
S.~Ghirlanda, 
O.~Hammarlid, 
M.~van~Hecke,
A.~Kozlov, 
G.~Lacorata,
A.~Martin-L\"of,
G.~Smirnov and
K.~{\.Z}yczkowski
for discussions. E.A. thanks 
Prof.~S.M.~Gama and
the Departamento do Matematica Aplicada of Universidade
do Porto for hospitality.
We thank the organizers of European Dynamics Days 1999
(Como June 20-23, 1999) and the
European Physical Society
Conference ``Applications of Physics to Financial
Analysis'' (Dublin July 14-17, 1999) for opportunities
to present this work, and the
Swedish Natural Research Council and the European Community for
financial support under grants
M-AA/FU/MA 01778-333~(E.A.) and
ERB4001GT962476~(P.M.-G.). 


\end{document}